\documentclass[prl,floatfix,twocolumn,showpacs,amsmath,amssymb,letter]{revtex4}

\usepackage{graphicx}
\usepackage{dcolumn}
\usepackage{bm}

\begin{document}

\def \cH{{\cal H }}
\def\nd{{^{\vphantom{\dagger}}}}
\def\yd{^\dagger}

\title{Charge density correlations in t-J ladders investigated
by the CORE method}

\author{Sylvain Capponi and Didier~Poilblanc}
\affiliation{Groupe de Physique Th\'eorique,
Laboratoire de Physique Quantique, UMR--CNRS 5626\\
Universit\'e Paul Sabatier, F-31062 Toulouse, France}
\homepage{http://w3-phystheo.ups-tlse.fr}

\date{\today}

\begin{abstract}

Using 4-site plaquette or rung basis decomposition, the CORE
method is applied to 2-leg and 4-leg t-J ladders and cylinders.
Resulting range-2 effective hamiltonians are studied numerically
on \emph{periodic rings} taking full advantage of the translation
symmetry as well as the drastic reduction of the Hilbert space. We
investigate the role of \emph{magnetic} and \emph{fermionic}
degrees of freedom. Spin gaps, pair binding energies and charge
correlations are computed and compared to available ED and DMRG
data for the full Hamiltonian. Strong evidences for \emph{short
range} diagonal stripe correlations are found in \emph{periodic}
4-leg t-J ladders.

\end{abstract}

\maketitle


Competition between superconducting correlations and
charge ordering 
has long been a challenge to numerical
computations~\cite{supercond,stripes}
of low-dimensional
strongly correlated electron
systems. Spin and hole-doped
ladders~\cite{DR96}
offer an ideal system to investigate the
cross-over between one to two dimensions. The 2-leg ladder for
example is known to exhibit a robust spin gap at and close to
half-filling as well as hole pair binding~\cite{DRS92,HPNSH95}.
Dominant power-law d$_{x^2-y^2}$-like pairing and $4k_F$ charge
density wave (CDW) correlations at small doping are characteristic
of a Luther-Emery (LE) liquid regime~\cite{LE95,LE96}. However, the
spin gap magnitude drops sharply as the number of legs is
increased, e.g. from $0.50\,J$ in the Heisenberg 2-leg ladder to
$0.190\,J$ in the Heisenberg 4-leg ladder~\cite{ladder_gaps}. Both
the increase of the magnetic correlation length (of the undoped
ladder) as well as the drastic reduction of the available ladder
length for increasing leg number restrict enormously the accuracy
of standard numerical techniques
like Exact Diagonalisation (ED) and Density Matrix
Renormalisation (DMRG) techniques. In addition, the
DMRG method is limited (in pratice) to Open Boundary Condition (OBC) in the
leg direction.

In this Rapid Communication, we use the Contractor Renormalisation
(CORE) method~\cite{weinstein96,altman01} to investigate 
hole-doped 2-leg and 4-leg ladders~\cite{capponi02}.
Our aim is to get further insights on the issue of pairing and
density correlations from the investigation of large enough
systems with \emph{Periodic Boundary Conditions} (PBC) in the
ladder direction. Such investigations are greatly needed to
complement available DMRG calculations using OBC. Our approach is
done in two steps; (i) first, using an appropriate partition into
small subsystems, we use the CORE method to construct an effective
hamiltonianx which integrates out quantum fluctuations at short
length scales; (ii) we use ED techniques (supplemented by 
finite size analysis) to compute various physical properties (pair
binding, spin gaps, etc...) and compare
them to those of the original model. The method provides new
results such as e.g. strong evidences for stripes correlations in
\emph{translationally invariant} 4-leg ladders.

\begin{figure}
\includegraphics[width=0.43\textwidth]{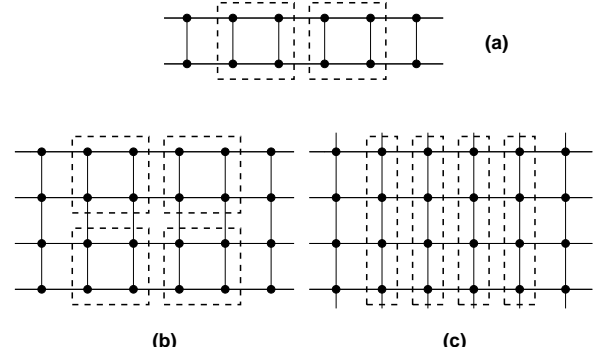}
\caption{\label{fig:lattices}
CORE decomposition in term of plaquette or rung sub-systems; (a)
2-leg ladder split in plaquettes; (b) 4-leg ladder split in $2\times 2$
plaquettes; (c) 4-leg cylinder split in 4-site rungs.
}
\end{figure}

We shall consider here a generic n-leg t-J ladder,
\begin{eqnarray}
\label{ham}
&{\cal H}&=J_{\rm leg}\sum_{i,a} \vec{S}_{i,a} \cdot \vec{S}_{i+1,a}
+J_{\rm rung} \sum_{i,a} \vec{S}_{i,a} \cdot \vec{S}_{i,a+1} \\
&+& t_{\rm leg} \sum_{i,a} (c_{i,a}^\dagger c_{i+1,a}+h.c.)
+t_{\rm rung} \sum_{i,a} (c_{i,a}^\dagger c_{i,a+1}+h.c.)\, ,
\nonumber
\end{eqnarray}
\noindent
where $c_{i,a}$ are projected hole fermionic operators.
Open (ladders) or closed systems (cylinders) along the rungs
with isotropic coupling, $t_{\rm leg}=t_{\rm rung}=1$
and $J_{\rm leg}=J_{\rm rung}=J$, will be of interest here.

In order to implement the CORE algorithm the ladders are
decomposed in small 4-site sub-units as shown in
Fig.~\ref{fig:lattices} whose $M$ low-energy states are kept to
define a reduced Hilbert space. The full hamiltonian~(\ref{ham})
is then diagonalised on $N$ connected units (with OBC) to retain
its $M^N$ low-energy states. These true eigenstates are then
projected on the reduced Hilbert space (tensorial product of the
$M$ states of each unit) and Gram-Schmidt
orthonormalized~\cite{weinstein96,altman01}. An effective
hamiltonian  containing N-body interactions with identical
low-energy spectrum can then be constructed in terms of the
reduced basis by a unitary transformation~\cite{truncated}. For
sake of simplicity, we shall restrict ourselves to the range-2
approximation($N=2$)~\cite{capponi02}.

At half-filling, retaining in each 4-site unit only the lowest
singlet and triplet states (4 states) gives excellent
results~\cite{capponi02}. Away from half-filling, the simplest
truncation, referred to as ``B'' approximation, is to
include, in addition, the lowest singlet hole pair on the 4-site
unit (of d-wave symmetry in the case of a plaquette). 
Formally, one can define 4 hard-core bosonic (plaquette or rung)
operators describing the 4 possible transitions from the singlet
half-filled GS (vacuum) to one component of the triplet state,
$t_{\alpha,i}$, or to the hole pair state, $b_i$
(Ref.~\onlinecite{riera02}). The effective B-hamiltonian $\cH^{B}$ 
can then be written as a sum of a simple bilinear kinetic term 
$\cH^{b}+\cH^{t}$ and a quartic interaction $\cH^{int}$
(Refs.~\onlinecite{altman01,capponi02}),
\begin{eqnarray}
 \cH^{b} &=& \epsilon_0+\epsilon_b \sum_i b^\dagger_{i}
b\nd_{i} - J_b \sum_{\langle ij\rangle} \left(
b^\dagger_{i} b\nd_{j}+\mbox{H.c.}\right)\\
\cH^{t} &=&  \epsilon_t \sum_{i\alpha  } t^\dagger_{ \alpha i}
t\nd_{ \alpha i} - \frac{J_t}{2} \sum_{\alpha \langle ij\rangle}
(t^\dagger_{\alpha i}t\nd_{\alpha j} + \mbox{H.c.})  \nonumber\\
&&~~- \frac{J_{tt}}{2} \sum_{\alpha \langle ij\rangle}
(t^\dagger_{\alpha i}t^\dagger_{\alpha j} + \mbox{H.c.})\, ,\\
\cH^{int} &=& V_b\sum_{\langle ij\rangle}n_{bi}n_{bj}
+\sum_{\langle ij\rangle}\big[V_0(t_it_j)^\dagger_0(t_it_j)\nd_0 \nonumber\\
&&+V_1(t_it_j)^\dagger_1(t_it_j)\nd_1
+V_2(t_it_j)^\dagger_2(t_it_j)\nd_2\big] \nonumber\\
&&-J_{bt}\sum_{\langle ij\rangle\alpha} ( b^\dagger_i b\nd_j
t^\dagger_{\alpha j}t\nd_{\alpha i}+ \mbox{h.c.} )
\nonumber\\
&&+V_{bt}\sum_{\langle ij\rangle\alpha} ( b^\dagger_i b\nd_i
t^\dagger_{\alpha j}t\nd_{\alpha j} + b^\dagger_j b\nd_j
t^\dagger_{\alpha i}t\nd_{\alpha i} )\,  ,
\label{eq:B_hamiltonian}
\end{eqnarray}
where $(t_i t_j)\yd_S$ creates two triplets on plaquettes $i$ and
$j$, which are coupled into total spin $S$. 
Such an effective
hamiltonian may serve for analytic and
numerical treatments. Its parameters listed for $J=0.35$ and
$J=0.5$ in table \ref{tab:tJ-par} are consistent with those
found for the Hubbard model\cite{altman01}. Although,
$\cH^B$ gives already a faithful description
of the physics of the original model,
a systematic improvement can be done by adding to the above
local basis 4 extra ``fermionic''
states corresponding
to the degenerate ($S_z=\pm 1/2$, even and odd chirality or
parity) single hole GS of the 4-site unit (hereafter referred to as ``BF''
approximation).

\begin{table}[ht]
\begin{tabular}{|c||c|c|c|c|c|c|}
\hline
$J$ & $\epsilon_0$ & $\epsilon_b$ & $\epsilon_t$ & $J_t$ &
$J_{tt}$ & $J_b$ \\
\hline 0.35 & -3.8895 & -3.5340 & 0.1379 & 0.2128 & 0.2319 & 0.2139 \\
0.5 & -5.5564 & -3.0919 & 0.1970 & 0.304 & 0.3112 & 0.2174 \\
0.35 & -3.5564 & -3.6579 & 0.4733 & -0.4836 & -0.4336 & 0.4855 \\
\hline \hline $J$& $J_{bt}$ & $V_b$ & $V_0$ & $V_1$ & $V_2$ & $V_{bt}$\\
\hline 0.35 &  -0.0709 &1.0345 & -0.1244 &
-0.0928 & 0.0412 & -0.3298\\
0.5 & -0.1044 & 0.8326 & -0.1777 & -0.1326 & 0.0588 & -0.3325 \\
0.35 & 0.2887 & 1.4164 & -0.2158 & -0.0202 & 0.0149 & -0.2489 \\
\hline
\end{tabular}
\vskip-0.4pc \caption{ Parameters of $\cH^{B}$ (in
units of $t$) computed for the t-J ladder
model using range-2 CORE with 2 plaquettes (row 1 and 2) or 2
4-site rungs (row 3).} \label{tab:tJ-par}
\end{table}

\begin{figure}
\includegraphics[angle=-90,width=0.43\textwidth]{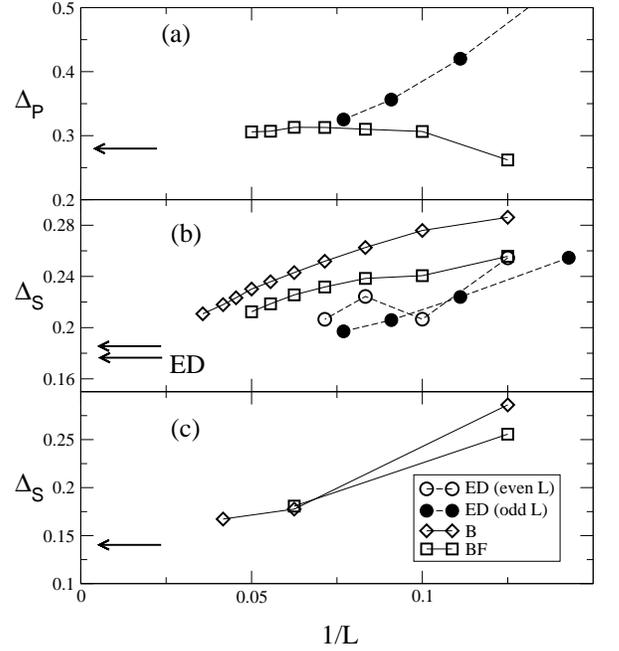}
\caption{\label{fig:2leg_ladder} Finite size scaling analysis vs
$1/L$ for a periodic \hbox{2-leg} $2\!\times\! L$ t-J ladder at
 $J=0.5$ using the decomposition of
Fig.~\protect\ref{fig:lattices}(a)
and the effective B- or BF- hamiltonians (as indicated on plot).
DMRG \& ED data for the original t-J ladder
are also shown for comparison. ED data obtained with odd ladder lengths
are averaged over boundary conditions
(see \protect\cite{poil00}). DMRG data and $L\rightarrow\infty$
ED extrapolations are shown by arrows.
(a) Pair-binding energy $\Delta_P$.
(b) Spin gap  of the
2-hole doped ladder.
(c) Spin gap of the $1/8$-doped ladder.
}
\end{figure}

\begin{figure}
\includegraphics[angle=-90,width=0.43\textwidth]{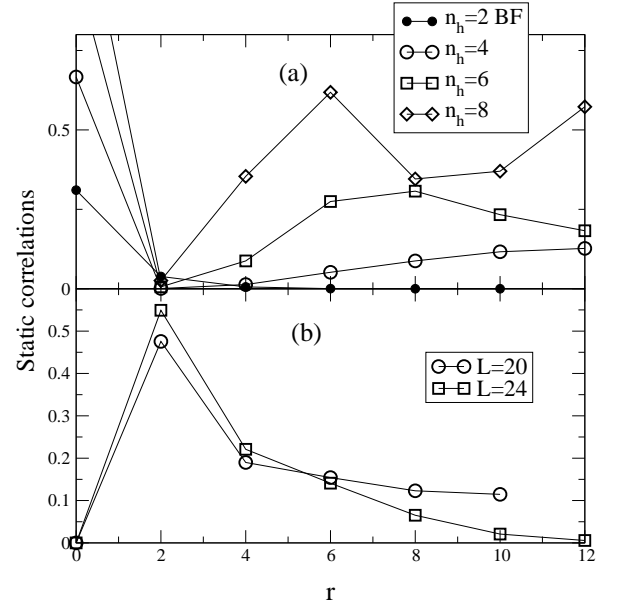}
\caption{\label{fig:corr_2leg} Correlations
as a function of distance (in units of the
original bond length)  up to $r=L/2$
in $2\!\times\! L$ t-J ladder at $J=0.5$.
(a) Plaquette charge density correlations for $L=24$.
The BF-hamiltonian (B-hamiltonian) is used for $n_h=2$ holes
(otherwise).
(b) Hole pair density-$S_z$
correlation in the lowest triplet excited state
of $2\!\times\! 20$ and $2\!\times\! 24$
2 hole-doped ladders using the B-hamiltonian.}
\end{figure}

\begin{figure}
\includegraphics[angle=-90,width=0.43\textwidth]{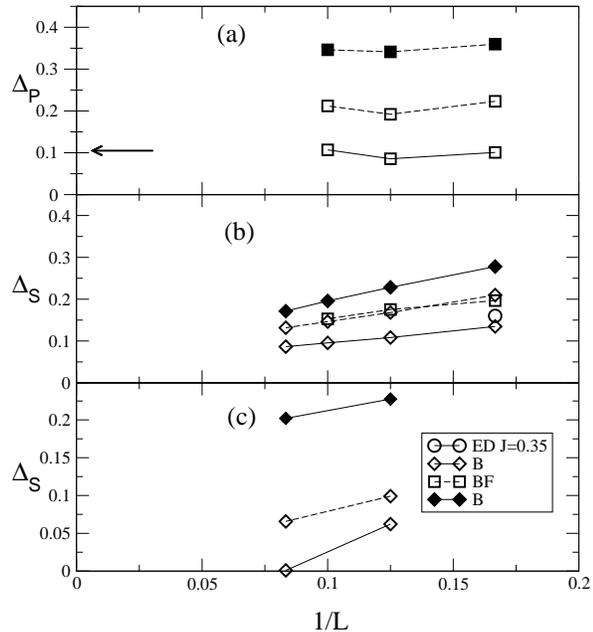}
\caption{\label{fig:4leg_ladder} Finite size scaling analysis
vs $1/L$ for a periodic \hbox{4-leg} $4\!\times\! L$
t-J ladder (open symbols) using the decomposition of
Fig.~\protect\ref{fig:lattices}(b)
for $J=0.35$ (full lines) and $J=0.5$ (dashed lines).
Data are also shown in the case of a periodic \emph{cylinder} (filled symbols)
for $J=0.35$ using the decomposition of Fig.~\protect\ref{fig:lattices}(c).
Symbols and notations similar to Figs.~\protect\ref{fig:2leg_ladder}(a-c).
(a) Pair-binding energy
$\Delta_P$. (b) Spin gap  of the 2-hole doped system. (c) Spin gap
of the $1/8$-doped system.}
\end{figure}

\begin{figure}
\includegraphics[angle=0,width=0.43\textwidth]{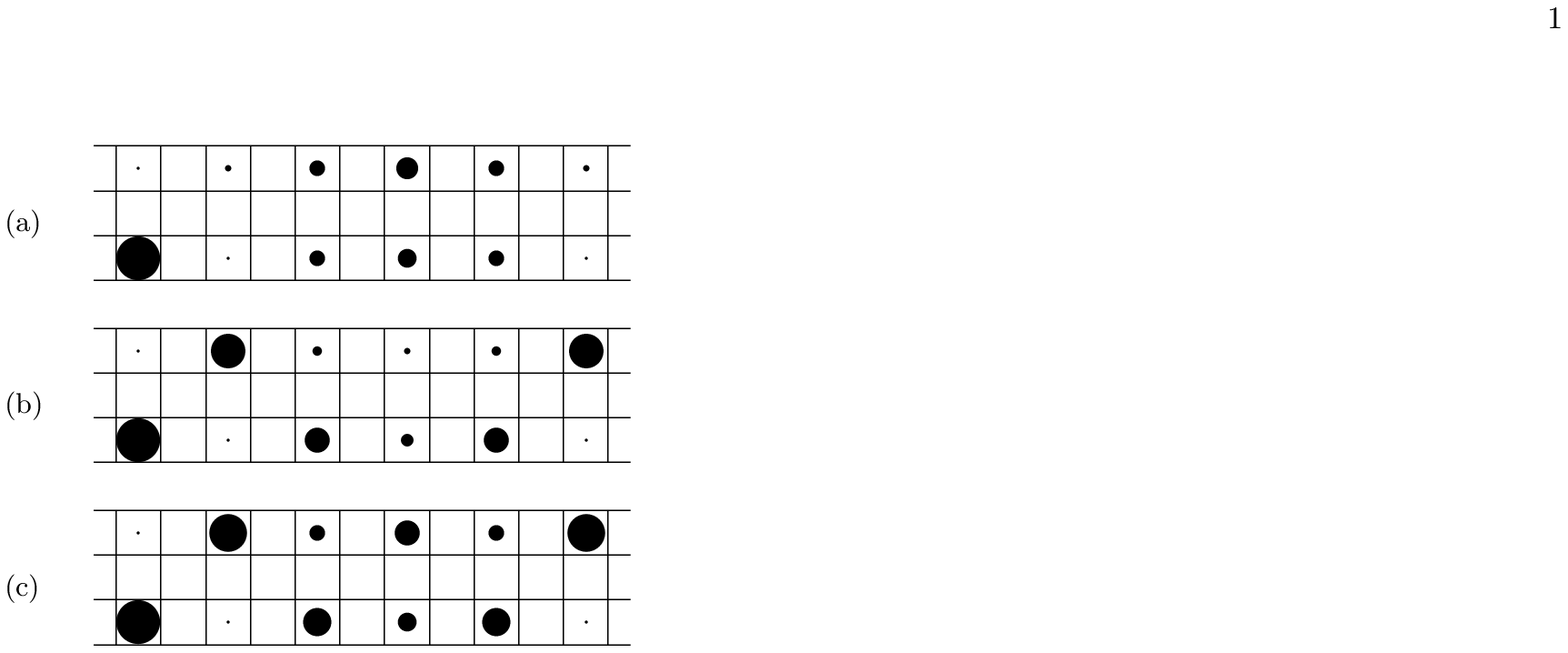}
\caption{Hole-pair density-density correlation on a $4 \times 12$ ladder
at $J/t=0.35$. PBC are used in the leg direction and
correlations are measured from the reference plaquette on the
lower left corner. From top to bottom, $n_h=4, 6, 8$. The surfaces of the
dots are proportional to the values of the correlations.}
\label{4hole_12}
\end{figure}

The 2-leg t-J ladder offers 
an ideal system to test the efficiency of the CORE method 
and the choice of the plaquette decomposition. As seen
from the behavior of the pair-binding energy
$\Delta_P=2E_0(n_h=1)-E_0(n_h=2)-E_0(n_h=0)$ plotted in
Fig.~\ref{fig:2leg_ladder}(a) and from the plaquette charge
density-density correlation in the two hole GS of the
BF-hamiltonian plotted in Fig~\ref{fig:corr_2leg}(a), pairs are
found to be strongly bound and localized almost on a single
plaquette. This confirms \emph{a posteriori} the relevance of CORE
and of the local basis. Furthermore,
the finite size scaling of the spin gap for a fixed number of
$n_h=2$ holes (see Fig.~\ref{fig:2leg_ladder}(b)) or at $1/8$ hole
density (Fig.~\ref{fig:2leg_ladder}(c)) gives gaps with 10-20\% accuracy 
in comparison to 
existing numerical data~\cite{poil00}. Due to
the small size of the hole pairs, accurate results are obtained
even when fermionic excitations are not
included. Note that the effective models lead to a smooth
finite size behavior, in contrast to the original t-J model where
``band-filling'' effects may lead to oscillatory behaviors as seen
in Figs.~\ref{fig:2leg_ladder}.

We point out the qualitative agreement between our
results and those of Siller et al.~\cite{siller01} who used a more
involved hard-core charged
boson model with longer range repulsive interactions (giving rise to a 
Luttinger liquid behavior), but neglected 
both fermionic and gapped triplet
excitations~\cite{siller01}. Our more systematic and general 
treatment using the B-hamiltonian gives a similar qualitative
picture as can be seen from the charge correlations shown in
Fig.\ref{fig:corr_2leg}(a); we observe the characteristic
4k$_F$-CDW spatial oscillations of the LE phase showing the same
number of maxima as the number of hole pairs. Let us emphasize
that this is also in agreement with DMRG
calculations~\cite{siller01}.
Our approach performed on finite homogeneous systems
is then complementary to the DMRG technique using OBC.

Although the agreement with the hard-core charged boson model is
qualitatively good, we believe that including magnetic triplet
excitations in the local basis is nevertheless important
to describe interplay between magnetic and
pairing correlations.
For example, it is known that the lowest triplet
excitation in a 2 hole doped (or very weakly doped)
t-J ladder consists of a hole pair-magnon boundstate~\cite{poil00}.
Indeed, the extrapolated value of
the spin gap in the presence of 2 holes (see
Fig.~\ref{fig:2leg_ladder}(b)) is lower than that of the undoped
ladder ($0.5J$) and than the hole pair binding
energy (shown in Fig.~\ref{fig:2leg_ladder}(a)).
Moreover, as seen in Fig.\ref{fig:corr_2leg}(b), the correlation
between the hole pair density and the plaquette $S_z$-component
clearly shows an enhancement at short distance~\cite{note1}.

We finish the investigation of the 2-leg ladder by using the 
effective Hamiltonian to calculate the Luttinger liquid parameter 
$K_\rho$ which governs the long distance power-law behavior 
of the charge correlations related to the unique massless charge mode.
Some values of $K_\rho$ obtained from the Drude weight $D$ and the 
compressibility $\kappa$~\cite{note2} as $K_\rho=\pi\sqrt{D\kappa/2}$ 
are listed in table~\ref{tab:Krho}.
In addition, we also list here the charge velocity $u_\rho$ obtained from
the relation $u_\rho=\pi D/K_\rho$ and which agrees within a few percents
to the values obtained directly from the linear dispersion of the charge mode. 
Note also that these values compare very well to existing ED~\cite{LE96} and
DMRG~\cite{siller01} data.

\begin{table}[ht]
\begin{tabular}{|c|c|c|c|c|c|c|c|}
\hline
doping & 14.3\%$^b$ & 12.5\%$^a$ & 10.7\%$^b$ & 8.3\%$^a$ & 7.1\%$^b$ & 4.2\%$^a$ &
3.6\%$^b$\\
\hline
$K_\rho$ & 0.559 & 0.602 & 0.668 & 0.753  & 0.798 & 0.914 & 0.920 \\
$u_\rho$ & 0.881 & 0.779 & 0.652 & 0.445 & 0.399 & 0.188 & 0.180 \\
\hline 
\end{tabular}
\vskip-0.4pc \caption{Parameters $K_\rho$ and  $u_\rho$ as a function of
doping computed on $2\times 24$ ($^a$) and $2\times 28$ ($^b$) ladder with B-hamiltonian and $J/t=0.5$.}
 \label{tab:Krho}
\end{table}
\vskip-1pc


We now turn to the investigation of the 4-leg t-J ladder (with OBC
along rungs) or cylinder (with PBC along rungs), for which the
best choices of unit decomposition are depicted in
Fig.~\protect\ref{fig:lattices}(b) and
Fig.~\protect\ref{fig:lattices}(c) respectively. Results for pair
binding energies and spin gaps are shown in
Fig.~\ref{fig:4leg_ladder}(a-c). Results for ladders and cylinders
are similar although the hole pair binding is much stronger in
cylinders where hole pairs are preferably formed on cross-sectional
plaquettes (periodic rungs)
rather than on ``surface'' plaquettes.
Generically we found that the pair binding energy is
larger than the spin gap of the undoped (Heisenberg) 
system
($0.190 J$ for the $4\times L$ ladder).
Therefore, the lowest triplet state in the 2-hole doped (or very lightly
doped) 4-leg ladder is similar to
a Heisenberg ladder magnon, which may be (or may not be) loosely bound to
a hole pair depending whether its excitation energy is
lower or equals the magnon energy of the undoped system.
Since the data shown on Fig.~\ref{fig:4leg_ladder}(b) are not fully
conclusive we have computed in addition the hole pair
density-$S_z$ correlation and found, as for the 2-leg ladder case,
an enhancement of the spin density on the neighboring sites of the
hole pair suggesting, indeed, the existence of
a hole pair-magnon boundstate.

Upon increasing doping, as seen from the hole-pair density-density
correlation shown in Fig.\ref{4hole_12}, we observe a clear
tendency of the hole pairs to align along the diagonal $(1,\pm 1)$
directions with a periodicity corresponding to one pair every two
plaquettes, a behavior also reported in DMRG
calculations\cite{white97,siller02} and reminiscent of the picture
of diagonal stripes.
Note that real space charge correlations are
fully consistent with the power law decay found
in the effective charge boson model~\cite{siller02}.

To conclude, the CORE method is a powerful
method to extract effective hamiltonians for strongly correlated
models. It allows numerical simulations on significantly larger
systems than those available for the original model. We show that
including charge \emph{and} spin bosonic excitations gives
reliable results as long as the hole pair binding energy is not
too small. Results for
the effective model of the 2-leg t-J ladder
are in excellent agreement with
known analytic and numerical data. Within the effective models
hole pair-triplet bound states form for both 2-leg and 4-leg
ladders, a key feature to be compared to $SO(5)$ phenomenological
theories\cite{zhang}. In addition, the method enables unbiased
(since calculated on translationnaly invariant clusters) analysis
of hole pair density correlations. While $4k_F$-CDW correlations
are found in 2-leg ladders, our computations provide clear
evidences in favor of short range diagonal stripes in 4-leg
ladders.

\acknowledgments We  thank IDRIS (Orsay) for allocation of CPU
time. We also acknowledge useful
discussions with E.~Altman, A.~Auerbach and S.C. Zhang.



\begin{thebibliography}{18}
\bibitem{note2} $D$ and $\kappa^{-1}$
are defined as second order corrections of the GS energy (per site) to a 
phase twist of the PBC and a change of the particle density respectively;
See e.g. \protect\cite{LE96}.
\end{thebibliography}
\end{document}